\documentstyle[marcin_conf_proc]{article}

\def\be{\begin{equation}}
\def\ee{\end{equation}}
\def\bea{\begin{eqnarray}}
\def\eea{\end{eqnarray}}

\begin{document}

\title{OPTICAL-IR SPECTRAL ENERGY DISTRIBUTIONS OF $z>2$ GALAXIES
\footnote{Based on observations made with the
NASA/ESA {\it Hubble Space Telescope} obtained at the Space Telescope
Science Institute, which is operated by the Association of
Universities for Research in Astronomy, Inc., under NASA contract NAS
5-26555.}
\footnote{Based on observations made at the Kitt Peak National Observatory,
National Optical Astronomy Observatories, which is operated by the
Association of Universities for Research in Astronomy, Inc.\ (AURA)
under cooperative agreement with the National Science Foundation.}
}

\author{ MARCIN SAWICKI and H.K.C. YEE}

\address{Department of Astronomy, University of Toronto, Toronto, M5S 3H8, Canada}
\address{e-mail: sawicki, hyee@astro.utoronto.ca
}


\maketitle\abstracts{
Broadband optical and IR spectral energy distributions are determined
for spectroscopically confirmed $z>2$ Lyman break objects in the
Hubble Deep Field.  These photometric data are compared to spectral
synthesis models which take into account the effects of metallicity
and of internal reddening due to dust.  It is found that, on average,
Lyman break objects are shrouded in enough dust (typically $E(B-V)
\approx 0.3$) to suppress their UV fluxes by a factor of $\sim 20$.
Furthermore, these objects are dominated by very young ($ < 0.2$ Gyr)
stellar populations, suggesting that star formation at high redshift
is episodic rather than continuous.  }

\section{Introduction}

A number of spectroscopically-confirmed $z > 2 $ Hubble Deep Field
(HDF) galaxies have been reported over the course of the last year
(Steidel et al.\ 1996b, Lowenthal et al.\ 1997).  The availability of
both the optical and infrared images of the HDF permits us to
construct broadband spectral energy distributions (SEDs) for these
galaxies.  
These SEDs sample the rest-UV and -optical, and hence contain
information about the galaxies' dust content and ages of stellar
populations.  Here, we outline the results of fitting these SEDs with
spectral synthesis models; full details of this work are described by
Sawicki \& Yee (1997).

\section{Data}

We used the Version~2 optical (Williams et al.\ 1996) and Version~1
Kitt Peak IRIM infrared (Dickinson et al.\ 1997) images of the Hubble Deep
Field.  
Object finding and photometry were
carried out using the PPP faint object photometry package (Yee, 1991),
while photometric calibrations were done using the zeropoints provided
by the STScI and Kitt Peak HDF teams.

We have measured broadband spectral energy distributions for the
seventeen $z>2$ objects listed in the catalogs of Steidel et al.\
(1996b), and Lowenthal et al.\ (1997).  (One of the Steidel et al.\
objects appears to consist of two distict subclumps which we have
treated as two objects at a common redshift).

\section{Model SEDs}

The observed broadband SEDs were fitted with 
GISSEL (Bruzual \& Charlot, 1996) multi-metallicity models, which had
been reddened with Calzetti's (1997) reddening recipe.
Time since the onset of star formation (``age'') and the amount of
reddening were free parameters.  The fits were done using the
$V_{606}$, $I_{814}$, $J$, $H$, and $K$, bands; the $U_{300}$ and
$B_{450}$ bands were not used so as to avoid the stochastic nature of
high-$z$ intergalactic attenuation (Madau, 1995) and the poorly-known
shape of the reddening law blueward of 1200 \AA.


\section{Results}

Figure 1a shows the ages and reddening parameters of the HDF $z>2$
galaxies, obtained by fitting $0.2Z_\odot$ model SEDs.  Note that the
vast majority of objects requires substantial amounts of internal
reddening, with $E(B-V) \approx 0.25$ being typical. Objects marked
with broken symbols are at $z>3$; due to bandpass shifting they have
only one filter above rest-4000 \AA\ and hence have poorly constrained
ages.

\begin{figure}
\includegraphics{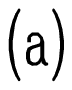}
\includegraphics{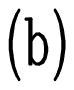}
\includegraphics{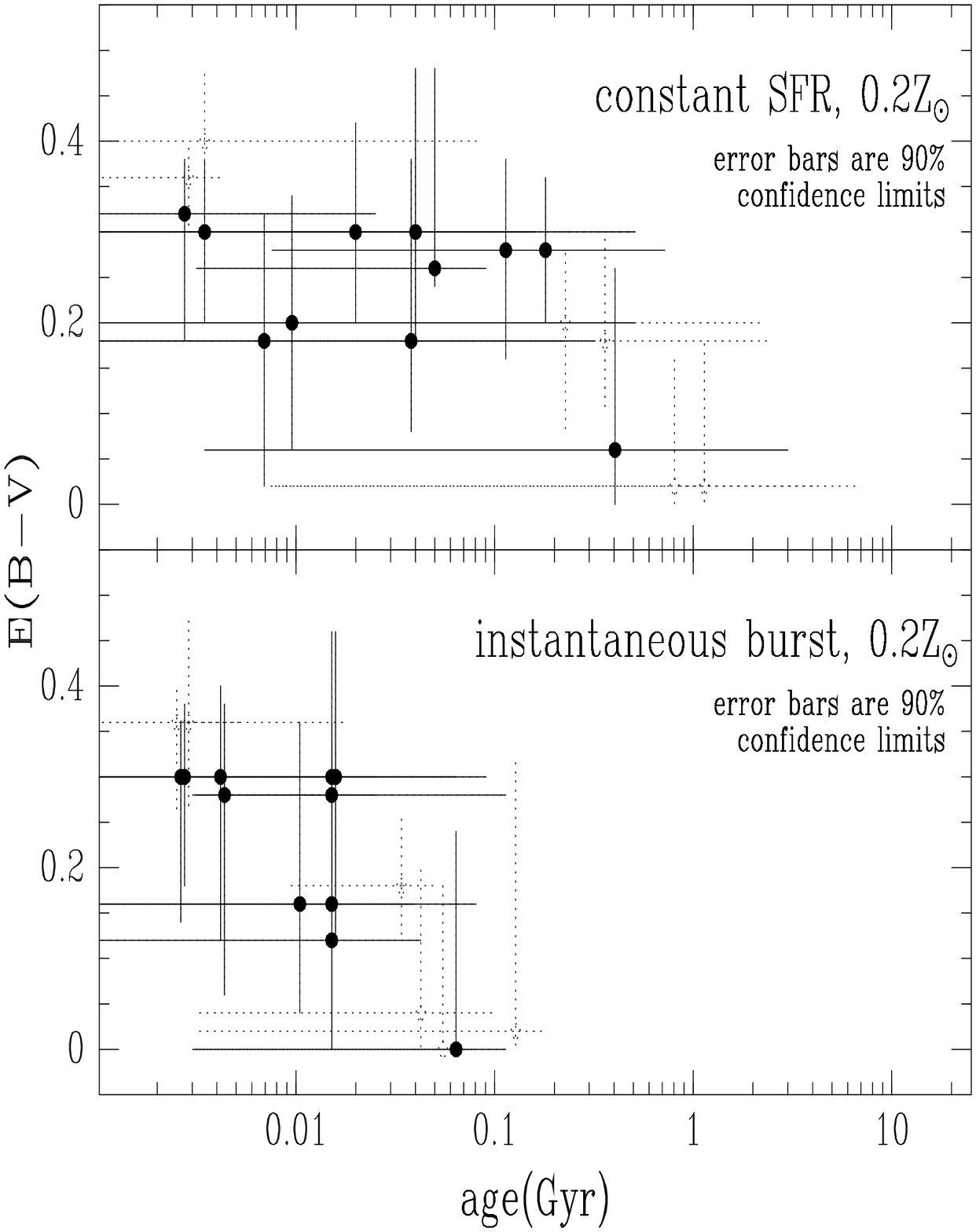}
\includegraphics{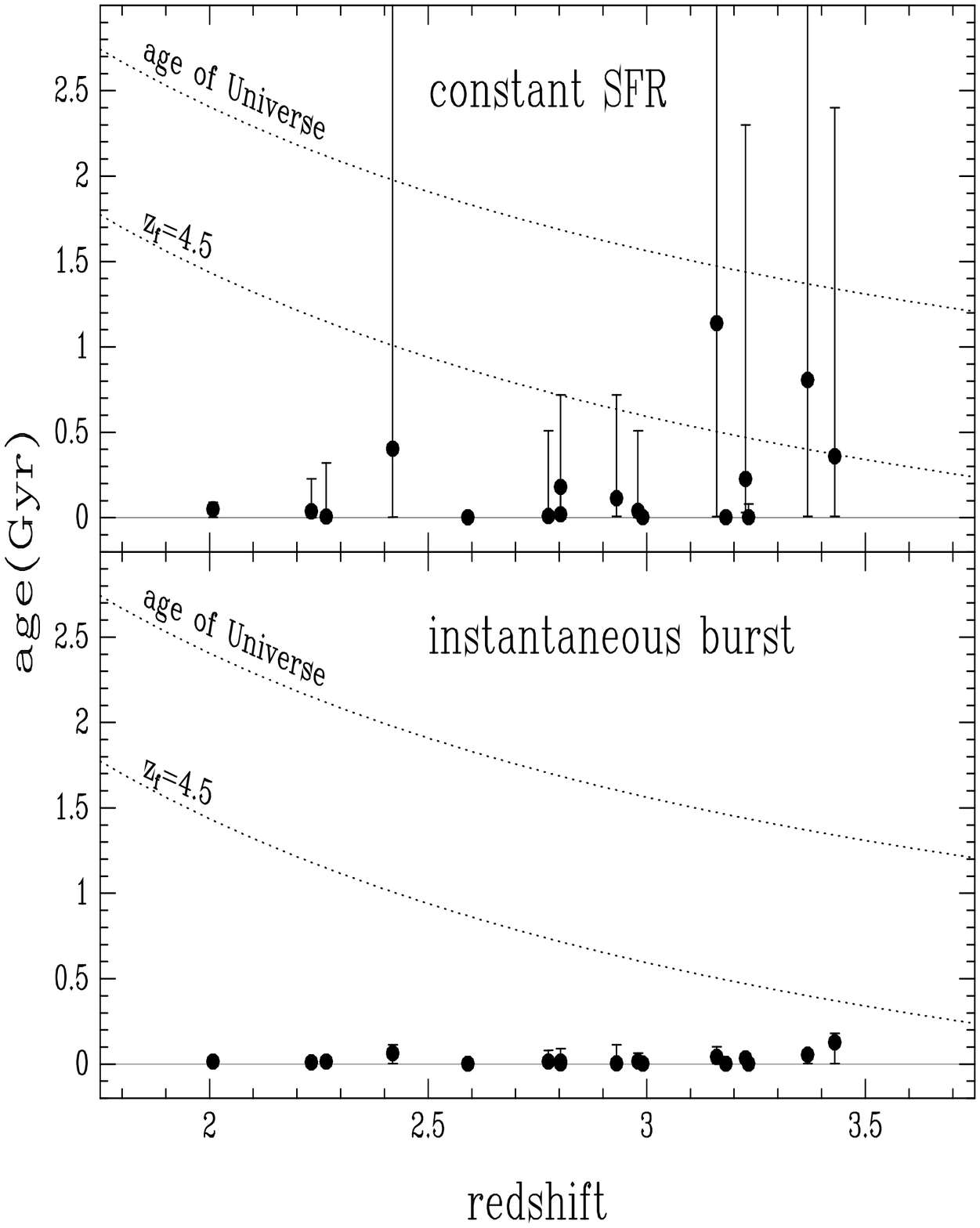}

\end{figure}

\begin{figure}
\vspace{8cm}
\caption{
{\bf (a)} Reddening and age of $z>2$ objects.  The top panel shows
the constant SFR fits and the bottom one is for the instantaneous
burst model.  Age means the time since the onset of star formation.
Galaxies at $z<3$ are shown as solid symbols while those at $z>3$ use
broken ones.  Error bars correspond to 90\% confidence limits.
{\bf (b)} The ages of domiant stellar populations of $z>2$ HDF objects
obtained by fitting $0.2Z_\odot$ model SEDs.  Error bars correspond to
90\% confidence limits.  The broken lines are the age of the universe
and the age of an object which formed at $z_f=4.5$ (both are for a
$\Omega=1$ universe whose present-day age is fixed at $t_0=12.5$ Gyr).
As is reflected in the sizes of uncertainties, objects at $z>3$ suffer
from poor coverage above $\sim 4000$~\AA\ and, consequently, have less
precise age estimates.  Stellar populations of the majority of $z>2$
HDF galaxies appear to have undergone recent ($t<0.2$ Gyr) episodes of
star formation.
}
\end{figure}

Figure 1b summarizes the ages of the best-fitting models.  The
instantaneous burst and constant SFR scenarios can be regarded as
limiting cases, with the actual star formation history, and hence age
of the dominant stellar population, falling somewhere in between these
two extremes.  Note that most of the high-$z$ HDF galaxies seem to
have undergone recent ($t<0.2$ Gyr) bursts of star formation

\section{Implications}

The $z>2$ HDF galaxies are best fitted with models containing
intrinsic reddening, with typical values of $E(B-V) \approx 0.25$.
Presence of dust has important implications for estimates of star
formation or metal ejection rates based on UV luminosities: at 1500
\AA, $E(B-V) = 0.3$ will attenuate rest-1500 \AA\ flux by a factor of
$\sim 20$ (see also Meurer et al.\ 1997; Meurer, this volume; Calzetti
1997).  UV-based estimates of star formation and metal ejection rates
at $z>2$ (e.g., Steidel et al.\ 1996a; Madau et al.\ 1996; Lowenthal
et al.\ 1997) have to be adjusted accordingly.

The dominant stellar populations of $z>2$ HDF galaxies appear to be
very young, consistent with having undergone a burst of star formation
within 0.2 Gyr prior to being observed.  The apparent deficit of older
(age $> 0.2$ Gyr) stellar populations at $z<3$ suggests that star
formation in high-redshift galaxies is episodic rather than
continuous.  While a continuously star-forming stellar population
would sustain its luminosity, a short-duration one will fade by $\sim
4$ magnitudes in 0.2 Gyr after the end of the burst, and hence will no
longer be detectable.  The HDF high-$z$ galaxies could therefore
represent a very select group of objects which had recently undergone
bursts of star formation and will fade out of the sample in a few
hundred Myr.

%
%

\end{document}